\documentclass[pra,twocolumn,aps,floatfix,showpacs,tightenlines,showpacs,superscriptaddress]{revtex4}
\usepackage{amssymb,amsmath,amsthm,amsbsy,epsfig,color,graphicx}

\newcommand{\beq}{\begin{eqnarray}}
\newcommand{\eeq}{\end{eqnarray}}
\newcommand{\be}{\begin{equation}}
\newcommand{\ee}{\end{equation}}

\begin{document}
\title{Genuine Multipartite Entanglement in the $XY$ Model}
\author{S. M. Giampaolo}
\affiliation{University of Vienna, Faculty of Physics, Boltzmanngasse 5, 1090 Vienna, Austria}
\author{B. C. Hiesmayr}
\affiliation{University of Vienna, Faculty of Physics, Boltzmanngasse 5, 1090 Vienna, Austria}



\begin{abstract}
We analyze the $XY$ model characterized by an anisotropy $\gamma$ in an external magnetic field $h$ with respect to its genuine multipartite
entanglement content (in the thermodynamic and finite size case). Despite its simplicity we show that the quantity ---detecting
genuine multipartite entanglement through permutation operators and being a lower bound on measures--- witnesses the presence of genuine multipartite entanglement for nearly all values of $\gamma$ and $h$. We further show that the phase transition and scaling properties are fully characterized by this multipartite quantity. Consequently, we provide a useful toolbox for other condensed matter systems, where bipartite entanglement measures are known to fail.
\end{abstract}


\maketitle

\section{Introduction}

Since the beginning of the century the analysis regarding entanglement properties of ground states of condensed matter systems has been on an equal footing with the Ginzburg-Landau approach to complex quantum many-body systems~\cite{AmicoFazioOsterlohVedral2008,EisertPlenioCramer2010,CalabreseCardyDoyon2009,Ladd2010,Osterloh2006}.
Further analyses have revealed the monotonic scaling of the von Neumann entanglement entropy in the ground state of spin models~\cite{Plenio1, Plenio2,VidalLatorreRicoKitaev2003,LatorreRicoVidal2004} -- the so-called {\em area-law} -- as well as profound relations to conformal field theory (CFT)~\cite{HolzheyLarsenWilczek1994,CalabreseCardy2004,CalabreseCardy2009}. The von Neumann entropy is the most celebrated member of the Renyi entanglement entropies. For example, the von Neumann is so important because of determining the scaling properties of numerical algorithms based on matrix product states~\cite{Shuchetal2008,Tagliacozzoetal2008,Pollmannetal2009,VerstraeteCirac2009}, but in terms of this research line it is particular useful because it lends itself to a useful tool to determine the continuous or discontinuous nature of a phase transition~\cite{ErcolessiEvangelistiFranchiniRavanini2011} and to estimate quasi-long-range order in low-dimensional systems~\cite{DalmonteErcolessiTaddia2011}.


Above all there is a growing awareness that the entanglement properties -- the whole is greater than the sum of its parts~\cite{Schroedinger} --
may provide the most fundamental characterization of quantum phases
of matter.  This is particularly true for cases concerning the presence of phases even in the
absence of local order parameters. Examples of this phenomenon
are represented by quantum phase transitions in which the ``ordered phase'' are characterized by a non-vanishing topological component of the
ground-state entanglement entropies~\cite{Balents2012,Kitaev2006,Levin2006}.

In this context, the entanglement spectrum and the topological component of the R\'enyi entropies
are being actively investigated in relation to Bose-Hubbard spin liquids~\cite{Isakov2011}, frustrated models on nontrivial lattice geometries~\cite{Vishwanath2011}, non-Abelian fractional Hall systems~\cite{Li2008}, and low-dimensional gapless models~\cite{ThomaleArovasBernevig2010,Hastings2012}.

In spite of recent exemplary findings concerning the analysis of bipartite entanglement, 
there still remains a number of unexplored avenues. Firstly, there is an undeniable rich structure to multipartite quantum correlations for which very little is understood. Secondly, since inseparability is known to be an NP-hard problem~\cite{Gruvits} and is devoid of computable necessary and sufficient criteria, this leaves sufficient-only criteria as the only way to tackle inseparability. However, note that many examples such as measurement based quantum computation~\cite{measurementbasedcomputer} or many other algorithms~(e.g. Ref.~\cite{BrussMacchiavello}) suggest that these properties are nonetheless important. In Refs.~\cite{Biseparabilitypaper1,Biseparabilitypaper2} a general framework to construct sufficient-only criteria was introduced; in particular, certain inequalities that have to be fulfilled in specific types of partially separable states.  If these inequalities are violated they prove partial inseparability, i.e. the presence of entanglement. This may in turn be used to show the extreme situation of a multipartite state that is not biseparable with respect to any bipartition, commonly referred to as a genuine multipartite entangled state (detailed definition in the next section).

There are only few papers dealing with partial separability of GME in many body system, e.g.~\cite{GuehneBriegel,GuehneToth,GabrielHiesmayr,Sanpera}. In Ref.~\cite{GabrielHiesmayr} the authors presented a novel method to explore genuine multipartite entanglement and partial inseparability in
many-body-systems by means of macroscopic observables only, such as energy. The idea is to compute the mean energy value of a given Hamiltonian
and to compare the value with the mean energy value minimized over all, e.g., bipartite states or, generally $k$-separable states. Since the mean
energy value is bound from below, the difference or gap between the energy value minimized over all k-separable states (choosing, e.g., $k=2$ equals GME) and the ground state energy values serves as a detection tool of partial separability. This introduced ``\textit{GME gap}" and ``\textit{k-entanglement
gap}", detects large areas of genuine multipartite entanglement and partial entanglement in typical many body states, however, it requires non-separable
ground states. In another recent paper~\cite{Sanpera} the genuine tripartite entanglement of the anisotropic $XXZ$ spin model was analyzed by applying proper constructed entanglement witnesses.

In this paper we shall follow a promising path of possibilities by applying the multipartite criterion introduced in Ref.~\cite{Biseparabilitypaper2}
directly to a subsystem of an one dimensional chain of spin-$\frac{1}{2}$ particles described by the $XY$ model in the thermodynamic case and the
finite size case.  In this respect, the paper takes the first steps to analyzing the properties of quantum multi-body systems via detection
tools for multipartite entanglement. We demonstrate that this method opens new avenues for successfully addressing properties of a condensed matter system such as quantum phase transitions or scaling properties. Note that few days after we put this paper on the net, two other related papers appeared on the ArXiv server. Namely, in Ref.~\cite{Campbell} the authors investigate quantum discord in the $XY$-model and in Ref.~\cite{Guehne3} the authors investigated also genuine multipartite entanglement in the $X$-model, however, different to this present paper their tool for detecting multipartite entanglement is based on a superset of the biseparable states.

The paper is organized as follows. In Sec.~\ref{MEB} we recall the definitions of partial separability and genuine multipartite entanglement. In Sec.~\ref{MODEL} we introduce the one dimensional spin-$\frac{1}{2}$ $XY$ model in an external magnetic field with periodic boundary
conditions. The principal characteristics are analyzed for both cases, i.e. for a finite size chain and in the thermodynamic limit. We have
chosen to focus our analysis on the $XY$ model since it represents a milestone in the field of the quantum entanglement relating to many body systems
(being one of the few models for which it is possible to evaluate analytically the reduced density matrix of the ground states regardless the
size of the system~\cite{LiebSchultzMattis,BarouchMcCoyDresded1970,BarouchMcCoy1971} and it turns out to be, e.g, a realistic model for helium
absorption on metallic surfaces~\cite{Domb} or can be realized with bosonic
atoms loaded in the p-band of an optical lattice in the Mott regime~\cite{Fernanda}). In Sec.~\ref{PANORAMIC} we analyze the presence of the genuine multipartite entanglement in the
$XY$ model paying particular attention to what happens close to the critical and factorization points.

\section{Detection of Multipartite Entanglement}
\label{MEB}

The theory of quantum entanglement describes the coherence properties of composite quantum systems. Indeed, the subtleties of composite quantum systems are such that there is no simple way to relate coherence properties of a single system to a larger composite system. However, one elementary method to characterize the coherence properties of composite systems is obtained in terms of the $k$-partite non-separability, i.e. the minimum number $k$ of entangled components necessary to describe an $n$-partite system.

\textbf{Definition: ``k-separability''} A pure $n$-partite quantum state $|\Psi_{(k-sep)}\rangle$ is said to be $k$-separable if and only if it
can be written as a product of $k$ states $\psi_i$ ($i=1,\dots k$) each one living on a different and non-overlapped Hilbert subspace
\begin{equation}
 \label{k-sep1}
 |\Psi_{(k-sep)}\rangle= |\psi_1\rangle \otimes |\psi_2\rangle \otimes \ldots |\psi_k\rangle\quad\textrm{with}\quad k\leq n \;.
\end{equation}
A mixed $n$-partite quantum state is said to be $k$-separable if and only if it has a decomposition into $k$-separable pure states:
\begin{eqnarray}
 \label{k-sep2}
 \rho_{(k-sep)}&=&\sum_i p_i |\Psi_{(k-sep)}^i\rangle\langle \Psi_{(k-sep)}^i|\nonumber\\
 &&\quad\textrm{with}\quad p_i\geq 0\quad\textrm{and}\quad \sum_i p_i=1\;.
\end{eqnarray}

From the above definition it immediately follows that if a state is $k$-separable, it is automatically also $k'$-separable for all $k'< k$. In particular, an $n$-partite state is called fully separable, if it is $n$-separable, whereas is called genuinely $n$-partite entangled, if and only if it is not biseparable ($2$-separable). In this interesting case all subsystems ``contribute'' to the entanglement. If neither of these is the case, the state is called partially entangled or partially separable.

While the above intuitive definition has been shown to provide a proper characterization of multipartite systems, it is far from straightforward
to determine whether or not a particular state is $k$-separable. This is immediately understood for $k$-separable mixed states since they may be
$k$-separable for various partitions. For example, a tripartite biseparable state can be given by
\begin{eqnarray}
&&\rho_{(biseparable)}=\nonumber\\
&&\sum p_i \rho_{AB}^i\otimes\rho_C^i+\sum q_j \rho_{AC}^j\otimes\rho_B^j+\sum r_k \rho_{BC}^k\otimes\rho_A^k\nonumber\\
&&\textrm{with}\quad p_i,q_j,r_k\geq 0\quad\textrm{and}\quad \sum_{i,j,k} p_i+q_j+r_k=1\;,
\end{eqnarray}
where $\rho_{xy}$ describe bipartite states. Consequently, the set of biseparable states are given by the convex hull of the states
that are separable for a fixed bipartition. From a physical point of view, the generation of fully separable or biseparable states does not
require the interaction of all parties. Moreover, from the quantum information theoretic and operational perspective genuine multipartite entangled
states allow for new applications such as quantum secret sharing (i.e. by distributing a secret over many different parties the genuine
multipartite entanglement assures security against eavesdropping or even unfair parties~\cite{SHH,HHB}).

The authors of Ref.~\cite{Biseparabilitypaper1,Biseparabilitypaper2} introduced a criterion that has to be satisfied for every $k$-separable
state. In the instance that it is violated, it serves as a test for multipartite entanglement. It is obtained by defining a set of permutation operators
$P_i$. Such operators act on two copies of an $n$-partite state, swapping the $i$-th subsystems between the two
copies, i.e.
\begin{eqnarray}
 \label{Pi-operator}
  P_i | \psi_{a_1,\cdots,a_n}\rangle \otimes |\psi_{b_1,\cdots,b_n}\rangle&=&|\psi_{a_1,\cdots,a_{i-1},b_{i},a_{i+1},a_n}\rangle \otimes
  \nonumber\\
  & & |\psi_{b_1,\cdots,b_{i-1},a_{i},b_{i+1},b_n}\rangle
\end{eqnarray}
where the $a_j$ and $b_j$ indicate the subsystems of the first and second copy of the state, respectively. It is evident that if the permuted
subsystem is separable from the rest of the state, the state is invariant under such a permutation. This observation is the key idea to constructing
very general separability criteria that are based on the following theorem proven in  Ref.~\cite{Biseparabilitypaper2}:``{\em
Every $k$-separable state $\rho$, regardless its purity, satisfies the following inequality
\begin{eqnarray}
 \label{criterion}
 &&I_k\!:=\!\\
 &&\sqrt{\!\langle\Phi|\rho^{\otimes 2}P_{tot}|\Phi\rangle\!}\!-\!\sum_{\{\alpha\}}\!\left(\! \prod_{i=1}^k
 \langle\Phi|P_{\alpha,i}^\dagger\rho^{\otimes 2}\! P_{\alpha,i}|\Phi\rangle\!\right)^{\frac{1}{2k}} \!\! \!\!\!\! \le \; 0\nonumber
\end{eqnarray}
for all fully separable states $|\Phi\rangle$, where $\rho^{\otimes 2}=\rho \otimes \rho$, the sum runs over all possible partitions $\alpha$
of the considered system into $k$ subsystems, the permutation operators $P_{\alpha,i}$ are the operators permuting the two copies of all
subsystems contained in the $i$-th subset of the partition $\alpha$ and \mbox{$P_{tot}=\prod_{i=1}^n P_{i}$} is the total permutation operator,
permuting the two copies.}''

The aim of this paper is to consider this criterion $I_k$ for $k=2$ applied to the ground states of the $XY$ model in order to prove the existence of
genuine multipartite entanglement (we will always consider the optimized value\footnote{The optimization is obtained by introducing $2\times 2$ unitary operators.}, i.e. optimized over all fully separable states $|\Phi\rangle$).
Importantly, this undertaking allows us to prove the existence if a set of genuine multipartite entangled states that do not posses the symmetries given by the vanishing commutators, $[P_{\alpha,i},\rho^{\otimes 2}]=0$. Obviously, non-violation of the criterion $I_2$  does not imply that the state of interest is not genuine multipartite entangled since the criterion $I_2$ is only necessary but not sufficient for
biseparability~\cite{Biseparabilitypaper2}. Let us remark that the obtained optimized value of $I_2$ are also lower bounds on multipartite
measures as shown in several Refs.~\cite{mconcurrence1,mconcurrence2,mconcurrence3,mconcurrence4,mconcurrence5,mconcurrence6}. In particular, these multipartite measures are based on the well-known concurrence~\cite{Wootters1998} applied to $4\times 4$ subspaces.

In the following we show that for most cases of the $XY$ model the criterion $I_2>0$ is helpful and for the few cases where the criterion fails
we discuss different strategies.

\section{The one dimensional $XY$ model}
\label{MODEL}

Having defined our machinery for the analysis of the multipartite entanglement we will proceed to introduce the well known XY model.
We consider a one dimensional chain of spin-$\frac{1}{2}$ particles, each localized at side $i$ and interacting with its nearest neighbor via an
interaction in $x$- and $y$-direction. Additionally, an external magnetic field in $z$-direction can be activated. We will pay particular attention to the class of translationally invariant states with periodic boundary condition. This one-dimensional ferromagnetic $XY$ spin model can be described by the following Hamiltonian
\begin{equation}\label{Hamiltonian1}
H_{xy}\! =\!-\sum_{i} \frac{1+\gamma}{2} \sigma_{i}^x \sigma_{i+1}^x\!+\! \frac{1-\gamma}{2}  \sigma_{i}^y \sigma_{i+1}^y - h\sum_{i}
\sigma_{i}^z \,,
\end{equation}
where $\sigma_{i}^\alpha$ $(\alpha \!= \! x,y,z)$ stands for the spin-$1/2$ Pauli operator on site $i$, $h$ is the strength of the external
transverse field, and $\gamma$ is the anisotropy of the nearest neighbor interaction in $x$- and $y$-direction, taking values in the interval
$[0,1]$. The extremes correspond, respectively, to energy-gapless isotropic $XY$ model ($\gamma=0$) and the Ising model ($\gamma=1$). Note that
the entanglement properties do not change when passing from the ferromagnetic to the anti-ferromagnetic $XY$ model ($H_{xy}\longrightarrow -H_{xy}$ and $h\longrightarrow -h$) since
only local unitary operators are needed to pass between these two cases.

Such models governed by the above defined Hamiltonian play a relevant role in the field of quantum statistical mechanics since, firstly, it is
one of the few models that can be solved analytically, and, secondly, it is a very good approximation to real condensed matter systems (e.g.,
for neutral atoms loaded on an optical lattice~\cite{Duan2003}), and, thirdly, it simulates a quantum circuit~\cite{Murg}.

The analytic solution
is obtained by applying the Jordan-Wigner transformations onto the spin operators obtaining pseudo-fermions~\cite{JordanWigner}. After these
transformations the system can be easily diagonalized. To obtain solutions for various finite sized systems and any value of the external magnetic
field $h$ we generalize the solutions obtained by Lieb {\em et al.}~\cite{LiebSchultzMattis}, whereas for the solutions in the thermodynamic
limit one can consult Refs.~\cite{BarouchMcCoyDresded1970,BarouchMcCoy1971}.

The model given by the Hamiltonian~(\ref{Hamiltonian1}) shows two important physical properties:
\begin{enumerate}
\item[1.] Regardless the value of anisotropy $\gamma$ and the size of the system, when the external magnetic field $h$ becomes equal to
$h_f \! =\! \sqrt{1\!-\!\gamma^2}$ the
system admits, among others, two fully factorized ground-states, i.e. global states that are products of single-site
states~\cite{GiampaoloAdessoIlluminati2008,GiampaoloAdessoIlluminati2009} (in  case of $\gamma \! = \!0$ into one single state). This case is
consequently named ``\textbf{factorization point}''.
\item[2.] In the thermodynamic limit for any value of the anisotropy $\gamma$ at the critical value of the magnetic field $h_{c}\! =\! 1$,
denoted  in the following as the ``\textbf{critical}'' point, the system undergoes to a \textit{quantum phase transition}: The ordered phase
($h<h_c$) possesses different properties in dependence of the anisotropy $\gamma$. For $\gamma>0$ the system is characterized by a twofold
degenerated ground-state space associated to a gapped energy spectrum; whereas for $\gamma=0$ the ground state space is not degenerated and
the energy spectrum is gapless. However, regardless the value of the anisotropy $\gamma$, in the paramagnetic phase $h > h_c$ the system is also
non-degenerated but the energy spectrum shows a gap between the ground state and the first excited state.
\end{enumerate}
The presence of a degeneracy in the ground-state-space forces us to make a choice among all possible ground-states of the system. To be
consistent between regions with and without degeneracy, we will always take into account that the ground-state preserves all the
symmetries of the Hamiltonian.

Moreover, we focus our interest on the presence of genuine multipartite entanglement between three adjacent spins, i.e.
position $i-1$ and $i$ and $i+1$. For such quantities, all the information needed is contained in
the reduced density matrix $\rho_3$ obtained from the ground-state wave function (after all the spins except those at positions $i-1$ and $i$ and $i+1$ have been traced out). Owning to the invariance of $\rho_3$ under spatial translation, the resulting matrix does not depend on the particular choice of $i$.

\section{Genuine Multipartite Entanglement in the $XY$ Heisenberg model}
\label{PANORAMIC}

We now present genuine multipartite entanglement features originating from the criterion $I_2$ for three adjacent spins of the
ground state, i.e. detecting the inseparability of this state into any two parts. We begin by considering the thermodynamic limit and follow this by
an examination of the finite sized systems. Then  we discuss scaling properties at the critical point $h=h_c=1$ and at the factorization point $h_f=\sqrt{1-\gamma^2}$ and last but not least we focus on the case of vanishing magnetic field $h=0$.

\subsection{Thermodynamic Limit}

\begin{figure}[t]
\includegraphics[width=8.cm]{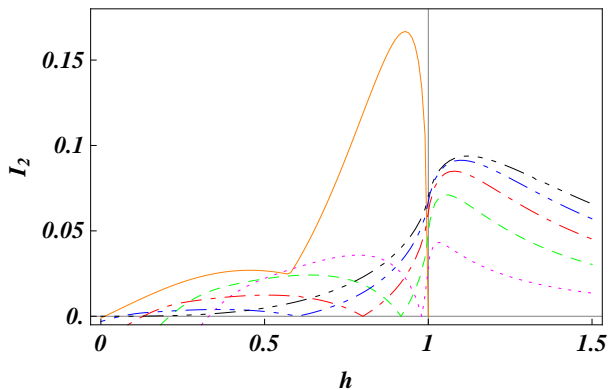}
\caption{(Color online) Comparison of the behavior of $I_2$ for the reduced density matrix $\rho_3$ as function of the external field $h$ for different values
of $\gamma \neq 0$ in the thermodynamic limit $N\rightarrow \infty$.
Orange solid line $\gamma=0$; magenta dotted line $\gamma=0.2$; green dashed line $\gamma=0.4$; red dot-dashed line $\gamma=0.6$;
blue dot-dot-dashed line $\gamma=0.8$ and black dot-dot-dot-dashed line $\gamma=1$ (Ising model). In the
paramagnetic phase $h>h_c=1$ the criterion $I_2$ detects genuine multipartite entanglement for all values of $\gamma$, except for the isotropic case
$\gamma=0$, which is known to be fully separable; whereas in the ferromagnetic phase $h<h_c=1$ only in the Ising case genuine multipartite
entanglement is detected for all values of the magnetic field.}
\label{Bisepthermolimit}
\end{figure}
In Fig.~\ref{Bisepthermolimit} the behavior of the criterion $I_2$ applied to $\rho_3$ in dependence of $\gamma$ and $h$ is summarized for the
thermodynamic limit case ($N \rightarrow \infty$). For all values of $\gamma \neq 0$ we observe that all the
curves follows a very similar behavior: at high values of the external magnetic field $h$ the criterion $I_2$ detects genuine multipartite entanglement, i.e.
$I_2>0$. Further lowering the external magnetic field $h$ the amount of the violation of the inequality increases until it reaches a maximum
close to the quantum critical point. Then it starts to decrease
(showing a diverge in the derivative exactly at $h=h_c \equiv 1$) until it vanishes at the factorization points $h_f=\sqrt{1-\gamma^2}$. Lowering
further the strength of the magnetic field $h<h_f$ the amount of the violation
increases again, reaching a second local maximum before dropping down and even becoming negative before the external field vanishes.
For $\gamma=0$ the behavior is very different reflecting the fact that the factorization field coincides with the critical case as well as the fact that for values greater than the critical field $h_c$ the ground state of the system is always the ferromagnetic state
$|\psi \rangle=\bigotimes_{i=1}^{N}|\uparrow_i \rangle$, i.e. a fully separable state.

\begin{figure}[t]
\includegraphics[width=8.cm]{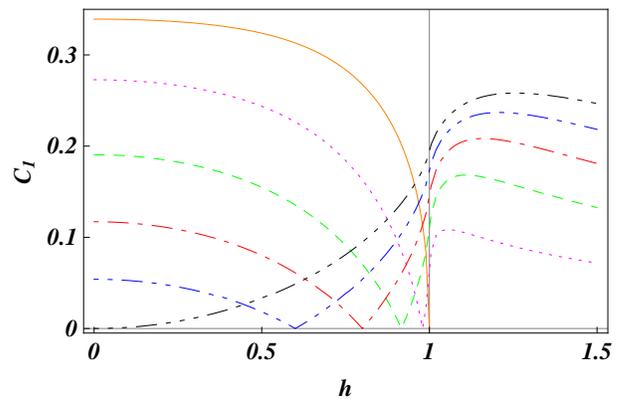}
\caption{(Color online) These graphs show the behavior of the nearest neighbor concurrence in dependence of the strength of the magnetic external
field $h$
for different value of $\gamma \neq 0$ in the thermodynamic limit $N\rightarrow \infty$: Orange solid line $\gamma=0$;
magenta dotted line $\gamma=0.2$; green dashed line $\gamma=0.4$; red dot-dashed line $\gamma=0.6$;
blue dot-dot-dashed line $\gamma=0.8$ and black dot-dot-dot-dashed line $\gamma=1$ (Ising model).}
\label{Concthermolimit}
\end{figure}
\begin{figure}[t]
\includegraphics[width=8.cm]{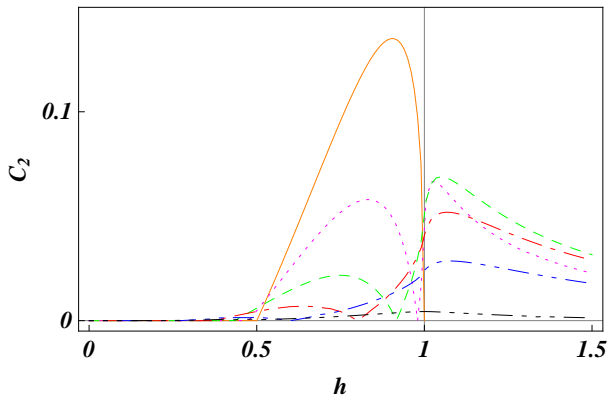}
\caption{(Color online) These graphs show the behavior of the next-nearest neighbor concurrence in dependence of the strength of
the magnetic external
field $h$ for different value of $\gamma \neq 0$ in the thermodynamic limit $N\rightarrow \infty$:
Orange solid line $\gamma=0$; magenta dotted line $\gamma=0.2$; green dashed line $\gamma=0.4$; red dot-dashed line $\gamma=0.6$;
blue dot-dot-dashed line $\gamma=0.8$ and black dot-dot-dot-dashed line $\gamma=1$ (Ising model).}
\label{Concnextthermolimit}
\end{figure}

It is natural to compare the behavior of $I_2$ for the three adjacent spins for the parity preserving ground state of
the $XY$ model in the thermodynamic limit with the behavior of the concurrence $C$~\cite{Wootters1998} for the two adjacent spins, i.e. the
reduced density matrix obtained tracing out the $(i-1)$-th spin of $\rho_3$. The result is summarized in Fig.~\ref{Concthermolimit}. In
Fig.~\ref{Concnextthermolimit} we have summarized the result of the concurrence computed for the next nearest spins, i.e. tracing over the $i$-th
spin.

Comparing the behavior of the quantity $I_2$ (Fig.~\ref{Bisepthermolimit}) with the one of the concurrence $C$ (Fig.~\ref{Concthermolimit})
we observe a
similar behavior above the critical point $h_c=1$. However, below the critical point the nearest neighbor entanglement increases and becomes
maximal at $h=0$. Different to the behavior of the nearest neighbor concurrence the next-nearest
neighbor concurrence becomes zero for small values of $h$ (see Fig.~\ref{Concnextthermolimit}). 
In this respect it is also interesting to investigate the question: what is the maximum bipartite entanglement in a translational invariant spin
chain? For finitely correlated qubit-chains the answer was found~\cite{HKN}, i.e. the concurrence obtains the maximal value of $C_{max}=0.434$.
We see that in our case
(Fig.~\ref{Concthermolimit}) that we are below this value.
In the Sec.~\ref{VANISHING} we investigate the case close to $h=0$ further.

%

\subsection{Finite size}

\begin{figure}[t]
\includegraphics[width=8.cm]{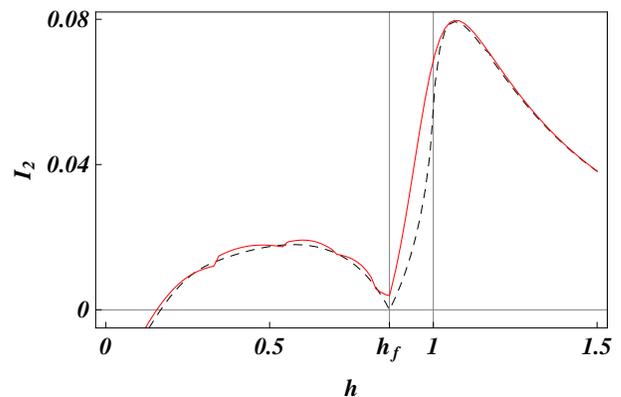}
\caption{(Color online) Comparison of the behavior of $I_2$ for the reduced density matrix obtained tracing out the degree of freedom of each
spins in the
system except three adjacent spins as function of the external field $h$ for $\gamma=0.5$ for different sizes $N$ of the chains;
black dashed line:
thermodynamic limit \mbox{$N \rightarrow \infty$}; red solid line: $N=14$.}
\label{Bisepfinitesize}
\end{figure}

Let us now compare the behavior of the the inseparability in the thermodynamic limit with finite size systems. In Fig.~\ref{Bisepfinitesize}
one observes three differences:

\begin{enumerate}
\item[1.] The first concerns the behavior of $I_2$ around the quantum critical point $h_c=1$. The function $I_2$ does not show (as expected)
the divergence of the derivative of the external magnetic field $h$. We discuss the finite size scaling in more details in Sec.~\ref{CRITICAL}.
\item[2.] The second concerns the factorization point $h_f=\sqrt{1-\gamma^2}$, here $I_2$ does not vanishes. That means that the genuine
multipartite entanglement decreases exponentially with the number of interacting spins.
This fact, at first sight, is very surprising because one is always driven to associate factorization with the vanishing of any type of
entanglement related quantity. However, as we discuss in more details in Sec.~\ref{FACTORIZING}, the factorization point is always, i.e.
regardless to the length of the chains, characterized by a two fold degenerate ground space~\cite{Blasone1,Blasone2,Rossignoli}.
\item[3.] The last difference is represented by a sort of oscillations in the region with $h<h_f$
that are associated to repetitive level crossing between
even and odd states~\cite{Blasone1,Blasone2,GiampaoloRenyi}. Such oscillations are also reflected in the behavior of bipartite entanglement~\cite{Ciliberti}.
\end{enumerate}

\subsection{Finite Size Scaling at Critical Point}
\label{CRITICAL}

\begin{figure}[t]
\includegraphics[width=8.cm]{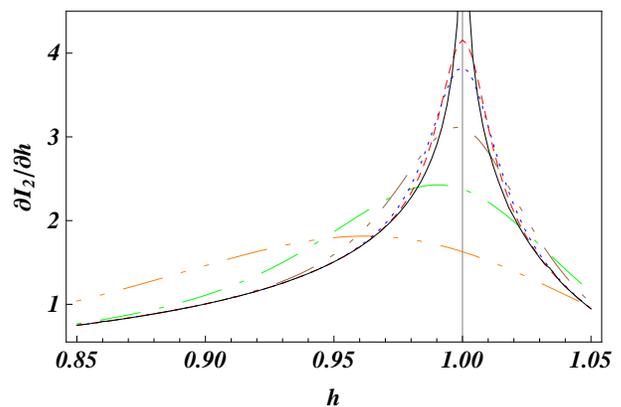}
\caption{(Color online) Behavior of the derivative $ \frac{\partial I_2}{\partial h}$ depending on $h$ in the case of $\gamma=1$ (Ising model)
for different
length (N): $N=20$ orange dot-dot-dashed line; $N=40$ green dot-dashed line; $N=80$ brown dot-dot-dot-dashed line;
$N=160$ blue dotted line; $N=220$ red dashed line; $N=\infty$ black solid line.}
\label{DerivBisepfinitesize}
\end{figure}

\begin{figure}[t]
\includegraphics[width=8.cm]{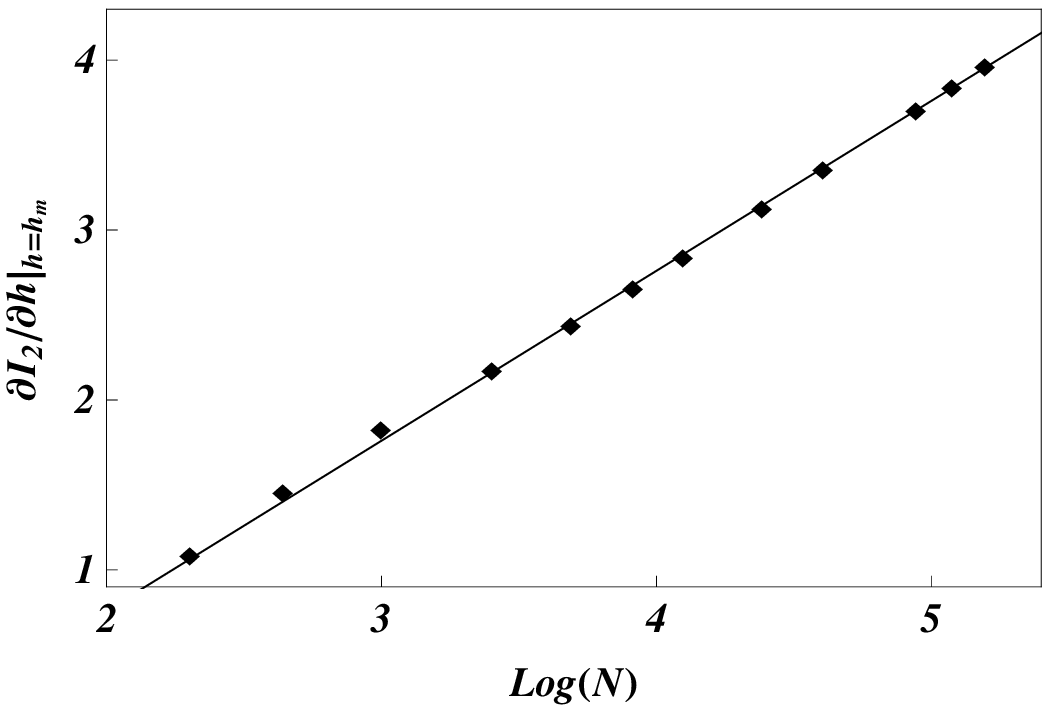}
\includegraphics[width=8.cm]{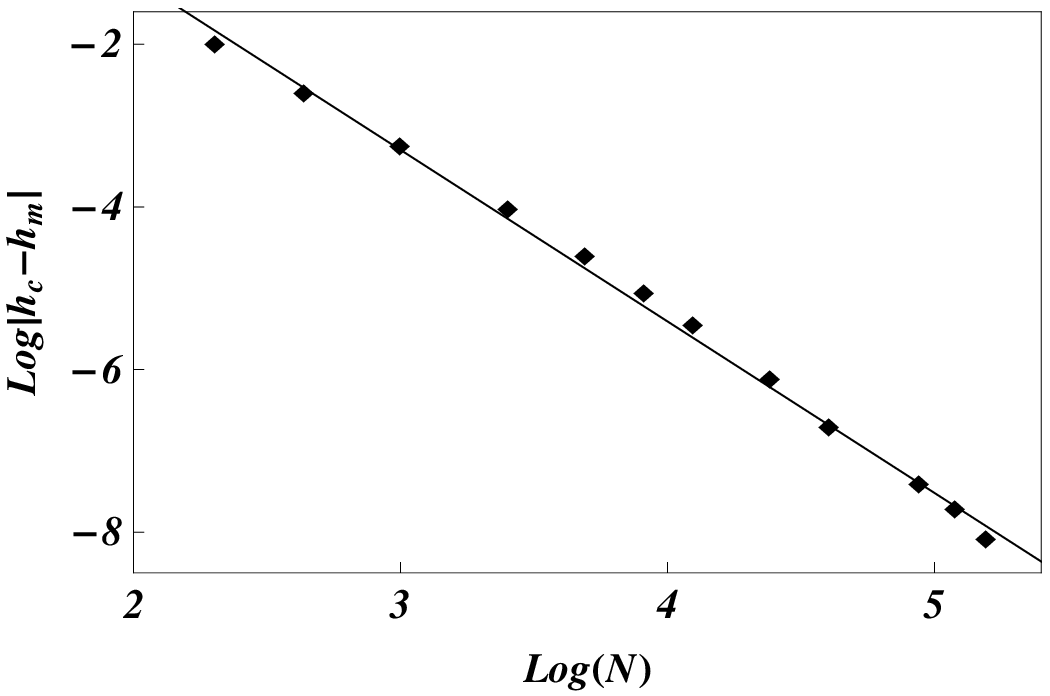}
\caption{ Upper panel ($\gamma=1$): Dependence of the maximum value of $\frac{\partial I_2}{\partial h}$ on the length of the chain $N$. Lower panel ($\gamma=1$):
Dependence of $h_{max}$, i.e. the value of the external field in which the maximum of $\frac{\partial I_2}{\partial h}$ is reached, on the size
$N$.}
\label{analysisofmax}
\end{figure}

In the thermodynamic limit at the critical point the derivative $\partial I_2/\partial h$ diverges. In the case of finite numbers of spins $N$
the derivative is presented in
Fig.~\ref{DerivBisepfinitesize}.
We observe by increasing the length of the chain and, hence, the system approaches the thermodynamic limit, the derivative shows an increasingly  pronounced maximum which becomes even more closer to the critical value $h=h_c \equiv 1$. Such qualitative descriptions can be quantified as
visualized in Fig.~\ref{analysisofmax}: In the upper panel we show how the maximum of the derivative of $I_2$ increase with
increasing length $N$, whereas in the lower panel we summarize how the external field -- for which the maximum of
$\frac{\partial I_2}{\partial h}$ is reached --, $h_{max}$, converges to the critical point $h=h_c \equiv 1$ for increasing $N$.
To be more precise the convergence of $h_{max}$ to $h_c\equiv 1$ is given by $1-h_{max} \simeq N^{-1.9} $, whereas
its maximum value diverges logarithmically with increasing system size $N$ as:
\begin{equation}
 \label{finitesizescaling1}
 \left. \frac{\partial I_2}{\partial h}\right|_{h=h_{max}}=0.78 \log N + const\;.
\end{equation}
This equation quantifies non-local correlations which are only due to genuine multipartite entanglement in the critical region.
\begin{figure}[t]
\includegraphics[width=8.cm]{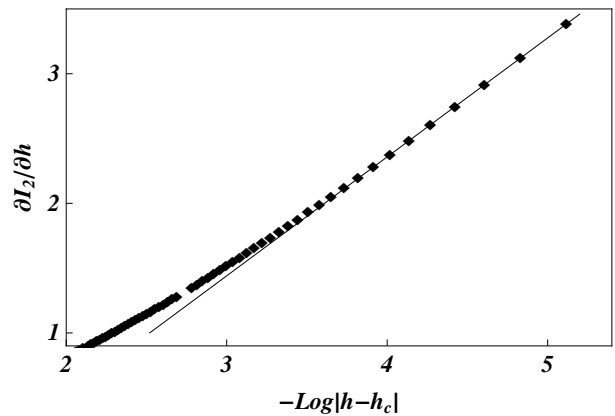}
\caption{Derivative of $I_2$ depending on the external magnetic field $h$ as a function of $-\log|h-h_c|$ in the thermodynamic limit.}
\label{derivativethermo}
\end{figure}
In the thermodynamic limit the derivative $ \frac{\partial I_2}{\partial h}$ diverges approaching the value
$h=h_c\equiv1$ (as shown in Fig.~\ref{derivativethermo}):
\begin{equation}
 \label{finitesizescaling2}
 \lim_{N\longrightarrow\infty}\frac{\partial I_2}{\partial h}=0.78 (- \log |h-h_c|) + const
\end{equation}
\begin{figure}[t]
\includegraphics[width=8.cm]{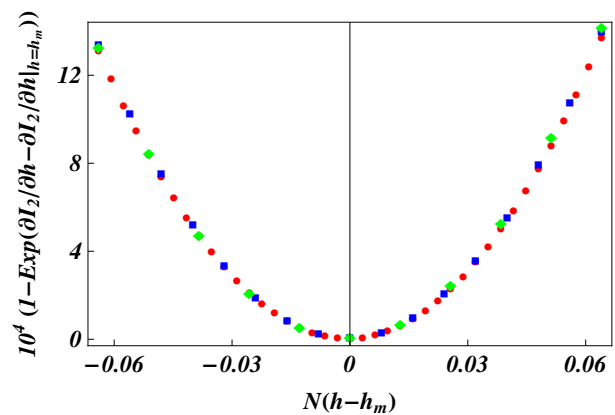}
\caption{(Color online) The universality (i.e. the critical properties depend only on the size of the system and the broken symmetry in the
ordered phase) of the (re-scaled) genuine multipartite entanglement is checked by plotting the finite size scaling with respect to $I_2$
in the case of $\gamma=1$. The red circles correspond to $N=14$ spins, the blue squares to $N=220$ spins and the green diamonds to $N=512$ spins.}
\label{testfinitesize}
\end{figure}
According to the scaling ansatz~\cite{Barber,Osterloh} in the case of logarithmic singularities, the ratio between the two prefactors of the
logarithm in Eq.~(\ref{finitesizescaling1}) and Eq.~(\ref{finitesizescaling2}) is the exponent $\nu$ that governs the divergence of the
correlation length $\xi\approx|\frac{1}{h}-\frac{1}{h_c}|^{-\nu}$. Consequently, arising from  Eq.~(\ref{finitesizescaling1}) and
Eq.~(\ref{finitesizescaling2}), it follows that $\nu=1$ (this result is consistent with the solution for the Ising model~\cite{Pfeuty}). This is
in agreement with the universality principle, a cornerstone of the theory of critical phenomena, i.e. the critical properties depend only on the
size of the system and the broken symmetry in the ordered phase.

By proper scaling~\cite{Barber} and taking into account the distance of the maximum from the critical point, it is possible to make all the data
for different $N$ collapse onto a single curve (see Fig.~\ref{testfinitesize}).
This figure contains the data for chain sizes ranging from $N=14$ up to $N=512$.

\begin{figure}[t]
\includegraphics[width=8.cm]{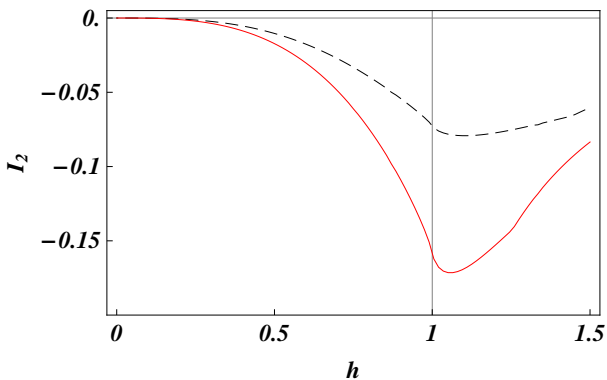}
\caption{(Color online) Comparison of the behavior of $I_2$ as function of the external field $h$ for reduced density matrix obtained
tracing out all the spins
except  $i-1,i,i+2$ (black dashed line), and $i-2,i,i+2$ (red solid line) for $\gamma =1 $  (Ising model) in the thermodynamic limit
$N\rightarrow \infty$.}
\label{Biseparability_as_func_of_r_gamma_1}
\end{figure}

One of the more surprising facts discovered by Fazio et al.~\cite{Osterloh} regarding the entanglement between pairs of spins is
that the range of entanglement remains finite, regardless of however close to the quantum critical point the range of the standard correlations diverges.  In Fig.~\ref{Biseparability_as_func_of_r_gamma_1} we plot the behavior of our genuine multipartite criterion for the Ising
model for nonadjacent spins, $i-1,i,i+2$ and $i-2,i,i+2$, respectively. Our criterion does not detect any genuine multipartite
entanglement. Even if this does not imply the absence of genuine multipartite entanglement at the critical point except in the case of three
adjacent spins, it strongly suggests that the range of entanglement is finite regardless the fact that it is bipartite or not.
\begin{figure}[t]
\includegraphics[width=8.cm]{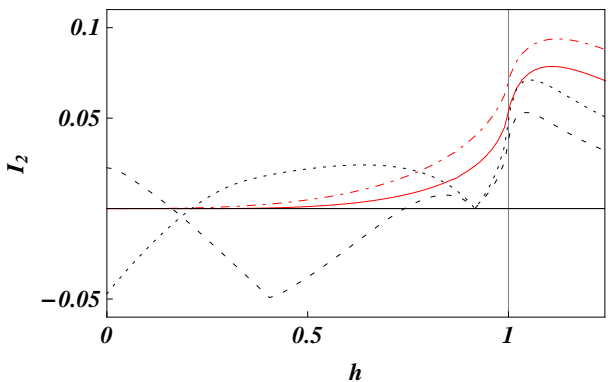}
\caption{(Color online) Comparison of the behavior of $I_2$ as function of the external field $h$ for reduced density matrix obtained tracing out
all the spins except three adjacent or four adjacent spins for different value of $\gamma$ in the thermodynamic limit $N\rightarrow \infty$.
Red dot-dashed line corresponds to the case of three adjacent spins and $\gamma=1$, whereas the red solid line stands for the case of four adjacent spins with the same $\gamma$. The black dotted line stands for the case of three adjacent spins and $\gamma=0.4$, whereas the black dashed line corresponds to the case of four adjacent spins with the same $\gamma$.}
\label{Bi_vs_triseparability_gamma_1}
\end{figure}

This is in contrast to the case where we compared genuine multipartite entanglement of three and four adjacent spins
(Fig.~\ref{Bi_vs_triseparability_gamma_1}), where we observe the same scaling behavior and a different behavior at $h$ close to zero. Let us remark here that it is not known how the detection ability of the criteria scales with the number of particles and with dimension.

Thus our results demonstrate that all the key ingredients of
the finite size scaling 
are present in the multipartite quantity. Our analysis shows that in the region of the quantum phase transition the bipartite and the multipartite entanglement of three
adjacent spins obey the same scaling behavior. 
In summary, the simple quantity $I_2$ turns out to be useful to analyze phase transitions and,
as such, offers a promising framework for fully characterizing phase transitions of quantum complex multi-body systems in terms of genuine
multipartite entanglement.


\subsection{Finite Size Scaling at the Factorization Point}
\label{FACTORIZING}

As we have noticed in Sec.~\ref{MODEL}, regardless the length of the chain, at the factorization point $h_f\equiv \sqrt{1-\gamma^2}$ the
system shows a degeneracy of the ground state~\cite{Blasone1,Blasone2} that is associated with the existence
of two factorized ground state (for $\gamma=0 $ in a single state),
of the form~\cite{GiampaoloAdessoIlluminati2008,GiampaoloAdessoIlluminati2009,Rossignoli}
\begin{equation}
 \label{factorizedstate1}
 | \Theta_{\pm} \rangle = \bigotimes_{i=1}^N \exp (\frac{i}{2} \theta_\pm \sigma_i^y) | \uparrow \rangle\;,
\end{equation}
where $\theta_\pm$ are fixed by the Hamiltonian parameters
\begin{equation}
 \label{factorizedstate2}
 \theta_\pm= \pm \arccos \left( \sqrt{\frac{1-\gamma}{1+\gamma}} \right)\;.
\end{equation}
However, for $\gamma \neq 0$ such states are not eigenstates with a fixed parity and hence are not the states on
which we are focusing our analysis, i.e. do not allow a coherent analyses. In such cases, the two eigenstates with fixed parity respect to
$\sigma_i^z$ can be constructed from the two factorized states, that represent a non orthogonal basis for the ground sub-space
\begin{eqnarray}
\label{factorizedstate3}
 | G_{even} \rangle & = &\frac{1}{Norm_+}\left( |\Theta_+\rangle + | \Theta_- \rangle \right)\,, \\
\label{factorizedstate4}
 | G_{odd} \rangle & = &\frac{1}{Norm_-}\left( |\Theta_+\rangle - | \Theta_- \rangle \right)\;,
\end{eqnarray}
where $Norm_+$ ($Norm_-$) is the normalization coefficient that ensures the normalization of the even (odd) ground state. As the two factorized ground states $| \Theta_{\pm} \rangle $ are, in general, not orthogonal to each other
\mbox{$(|\langle \Theta_+ |\Theta_- \rangle|^2=\cos^{2N}\theta)$}, the required normalization coefficients are non-trivial and can be deduced as
\begin{equation}
 \label{factorizedstate5}
  Norm_\pm= \sqrt{2 (1 \pm \cos^N \theta_i)}\;.
\end{equation}

\begin{figure}[t]
\includegraphics[width=8.cm]{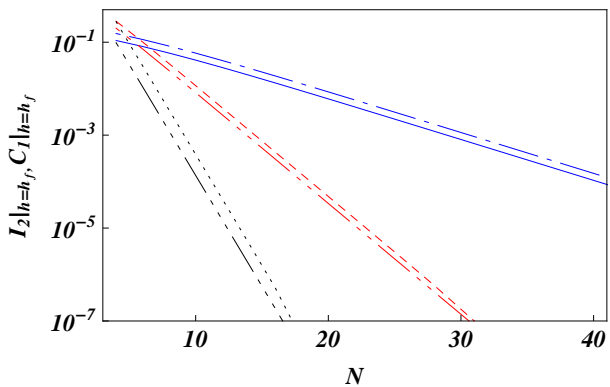}
\caption{(Color online) Comparison between the behavior of $I_2$, and the behavior of the nearest neighbor concurrence $C_1$ at the factorization point as function of the size of the chain $N$ for different values of $\gamma$:
$\gamma=0.2$ (blue) solid line $C_1$ and dot-dashed line $I_2$;  $\gamma=0.5$ (red) dashed line $C_1$ and dot-dot-dashed line $I_2$;
$\gamma=0.8$ (black) dotted line $C_1$ and dot-dot-dot-dashed line $I_2$.}
\label{bisepatfactorization}
\end{figure}

\begin{figure}[t]
\includegraphics[width=8.cm]{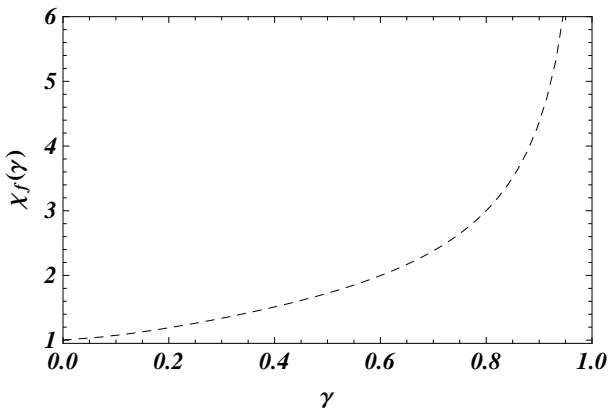}
\caption{Behavior of $\chi_f$ as function of $\gamma$.}
\label{argofexp}
\end{figure}

The two factorized ground state are fully separable and hence, for them, any quantity related to the entanglement must vanishes. However,
due to the non-orthogonality of the even and odd ground states the resulting state can be entangled. In Fig.~\ref{bisepatfactorization} we show
the detailed behavior of $I_2$ and the nearest neighbor concurrence $C_1$ at factorization point as function of $N$ and different $\gamma$'s.
We observe that increasing the size of the chain both entanglement quantities decrease with an exponential law proportional to
$\chi_f^{-N}$, where $\chi_f$ is a function of $\gamma$ that is plotted
in Fig.~\ref{argofexp} (that depends on the inverse of the energy gap between the ground state and the first excited states~\cite{Sachdev}).
Increasing $\gamma$ the function $\chi_f(\gamma)$ increases resulting in a faster and faster exponential decay. For the Ising model
$\gamma=1$ it diverges signaling that, in such a case, both $I_2$ and the concurrence $C_1$ of the even and odd
ground states vanishes for any $N$. In particular, the ground state is exactly the GHZ state, hence, tracing out any part of the system will reveal the disappearance of all entanglement.

For the opposite limit, i.e. when $\gamma=0$ for any $h\ge1$ and any $N$, the system admits as unique ground state
$|\psi \rangle=\bigotimes_{i=1}^{N}|\uparrow_i \rangle$ and
hence all entanglement vanishes. For all other cases, i.e. for $0<\gamma<1$, we find that due to the non-orthogonality of $|\Theta_+\rangle$ and
$|\Theta_-\rangle$ multipartite entanglement disappears only in the thermodynamic limit. In this case
$\rho_3$ can be regarded as an equally weighted mixture of the two fully separable states, $|\Theta_+\rangle$ and  $|\Theta_-\rangle$.

\subsection{Analysis of Multipartite Entanglement at $h=0$}
\label{VANISHING}

As evident from Fig.~\ref{Bisepthermolimit} and Fig.~\ref{Bisepfinitesize} and regardless the size of the
system, the criterion $I_2$ becomes negative at sufficiently small value of the external field $h$ for $0<\gamma<1$. On the other hand, for $\gamma=0,1$
it vanishes exactly only at $h=0$. Since the positivity of $I_2$ is only a necessary but not sufficient criterion -- based on permutation of all
possible bipartitions of the two-copy state -- we cannot conclude that the genuine multipartite entanglement vanishes for small $h$. However, by approaching the problem numerically via Monte Carlo techniques, we may be able to exploit various known symmetries. Taking a Monte Carlo approach and defining $\sigma_3$ as a generic three spin state obtained as a mixture of fully separable and biseparable state we sought to
minimize the Hilbert-Schmidt distance~\cite{Bhatia} defined by
\begin{eqnarray}
  D(\rho_3,\sigma_3)&=&\frac{1}{2} \mathrm{Tr}\|\rho_3-\sigma_3 \|^2 \\
  &=& \frac{1}{2} \left[ \mathrm{Tr}(\rho_3^2) +\mathrm{Tr}(\sigma_3^2)-2 \mathrm{Tr}(\rho_3 \sigma_3)\right]\;. \nonumber
\end{eqnarray}
In all our simulations we choose as starting point a mixture composed of two pure fully separable state with the same weight:
\mbox{$|a_+\rangle\!=\!|+\rangle\! \otimes \! |+\rangle \! \otimes \! |+\rangle $} and
\mbox{$|a_-\rangle\!=\!|-\rangle\! \otimes \! |-\rangle \! \otimes \! |-\rangle $} where $|+\rangle $
($|-\rangle $) is the eigenvalues of $\sigma_x$ with  eigenvalues $1$ ($-1$). We based this choice on the fact that such states are equal to $\rho_3$ for $\gamma=1$ (quantum Ising model), regardless of the chain length. Therefore $\rho_3$ for the quantum Ising model is always, i.e.
for any $N$, fully separable. Our algorithm starts by defining a mixture of $6$ fully separable states and $12$ biseparable states
($4$ states with entanglement between spin $i-1$ and spin $i$, $4$ with entanglement between spin $i-1$ and $i+1$, and $4$ with entanglement between $i$ and $i+1$).
Then the distance is computed and compared to the distance for variations of the initial state. If the distance is smaller, the varied state
becomes the new initial one, otherwise the initial state is kept, however, only with probability equal to the negative exponent of the difference of the two distances.

\begin{figure}
\includegraphics[width=8.cm]{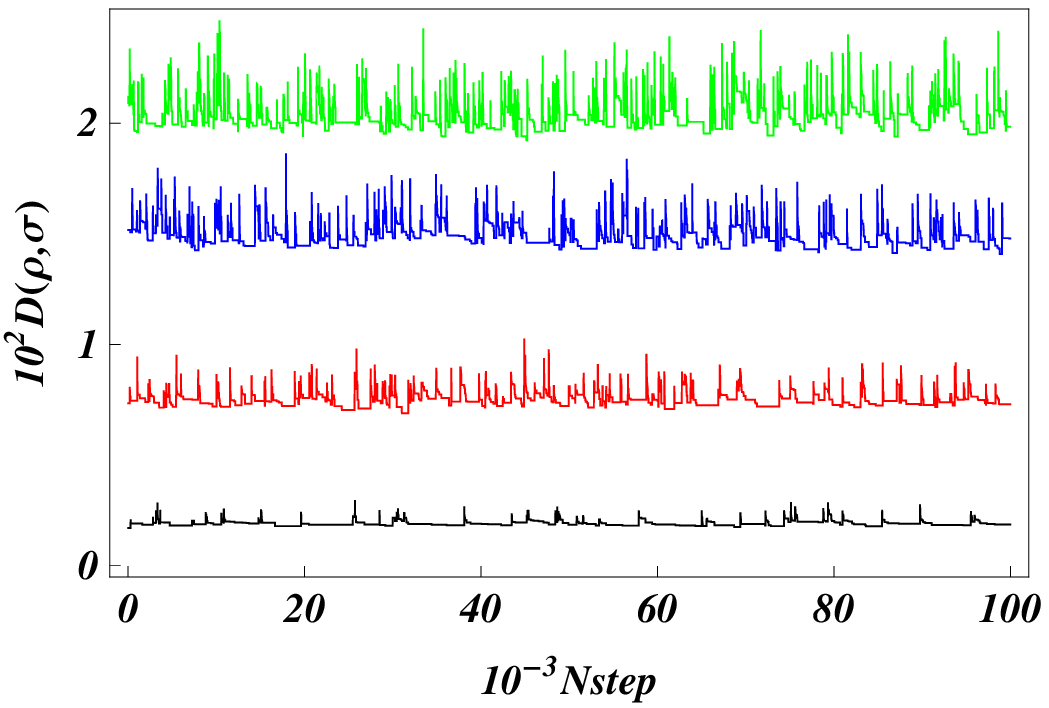}
\includegraphics[width=8.cm]{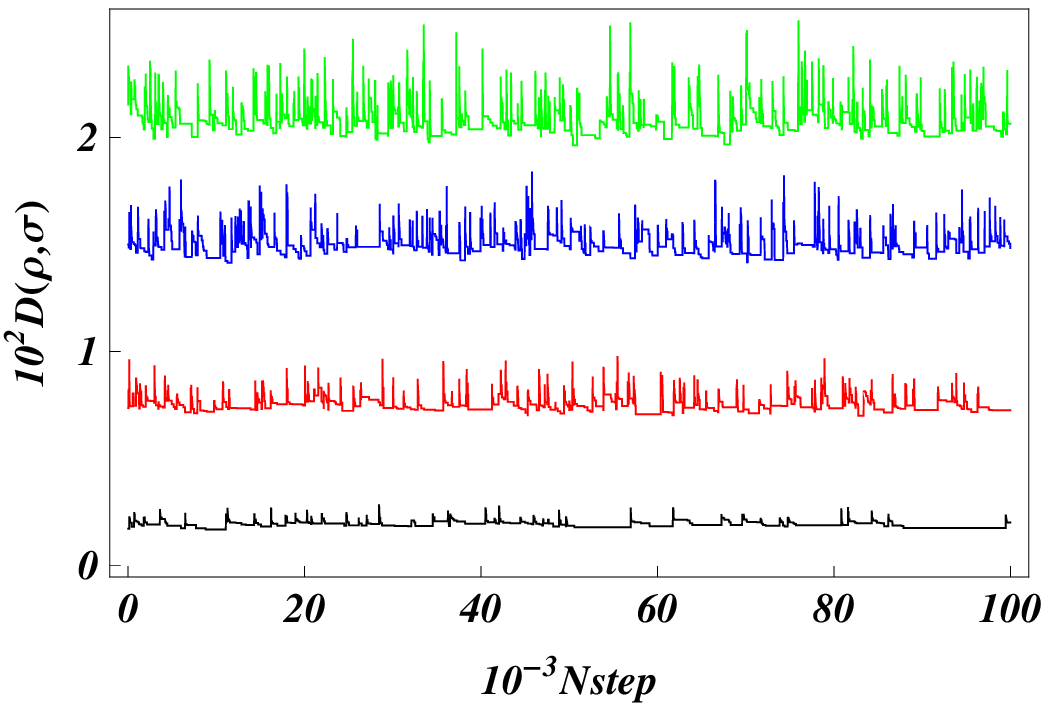}
\caption{(Color online) Behavior of the (re-scaled) Hilbert-Schmidt distance $D(\rho_3,\sigma_3)$ as a function of the steps $Nstep$ in the
Montecarlo simulations  for different values of $\gamma$ at $h=0$
for a chain of infinite length (upper panel) and for a chain of $N=20$ spin (lower panel).
From the bottom to the top: black lines $\gamma=0.75$; red lines $\gamma=0.5$; blue lines $\gamma=0.25$; green lines $\gamma=0$.}
\label{montecarlo}
\end{figure}

The results of our Montecarlo numerical simulation are summarized in Fig.~\ref{montecarlo} where the upper (lower) panel shows the
behavior of the distance $D(\rho_3,\sigma_3)$ as a function of the step of the Montecarlo algorithm for the system in the thermodynamic limit
(of $N=20$ spins) for different values of $0\le \gamma<1$.
All samples follow the same behavior: after a very fast decrease the algorithm reaches a threshold (that depends strongly on
$\gamma$ and seems also to depend only weakly on $N$) that seems impossible to be overcome. Even if these numerical simulations do not
represent a proof of the presence of the multipartite entanglement in the $XY$ model, the lack of convergence in our algorithm is a strong
suggestion; in agreement with the results presented in Ref.~\cite{GiampaoloRenyi}. We also wish to draw the readers attention to Fig.~\ref{Bi_vs_triseparability_gamma_1} where we present the criterion $I_2$ applied to four adjacent spins and observe genuine multipartite entanglement for vanishing magnetic fields.

\section{Conclusions}

In conclusion, in this paper we have analyzed the presence of genuine multipartite entanglement in the ground-state of the $XY$ model.
Despite its simplicity the $XY$ model being characterized by the anisotropy parameter $\gamma$ and an external magnetic field $h$ possesses
genuine multipartite entanglement for almost all values of $\gamma$ and $h$. From a physical point of view this means that the entanglement present in a subset of a spin chain does require interaction of all particles. Moreover, by analyzing in detail the finite size case and the thermodynamic
case we proved that the considered criterion detecting genuine multipartite entanglement -- being also a good bound on multipartite measures -- is a proper quantity to explore all the features of phase transitions and scaling properties.

Consequently, further analyses of distinct types of genuine multipartite entanglement or different multipartite criteria open up a promising
toolbox to characterize properties of condensed matter systems. In particular, our approach may turn out to be extremely useful for systems that
show topological order; in this case the bipartite entanglement is not able to provide a complete characterization of the ground state
properties~\cite{Balents2012,Kitaev2006,Levin2006}. Other systems where our approach applies are frustrated quantum spin
models~\cite{frust1,frust2};  here some preliminary results suggest that the transition between a saturated ground-state and a non-saturated
ground state is due to an abrupt change of the relative weight in the multipartite entanglement properties.

\section*{Acknowledgments}
We want to thank Colin Wilmott and the referees for many useful suggestions to improve the paper. SMG and BCH acknowledge gratefully the Austrian Science Fund (FWF-P23627-N16).

\end{document}